\documentclass[12pt,epsf,amstex]{article}
\usepackage [dvips]{graphicx}
\usepackage{amsmath}
\usepackage{amssymb}
\usepackage{epsfig}
\usepackage{verbatim}

\addtocounter{secnumdepth}{1}
\font\capfont=cmbx12 at 50 pt 
\newbox\capbox \newcount\capl \def\a{A}
\def\docappar{\medbreak\noindent\setbox\capbox\hbox{%
\capfont\a\hskip0.15em}\hangindent=\wd\capbox%
\capl=\ht\capbox\divide\capl by\baselineskip\advance\capl by1%
\hangafter=-\capl%
\hbox{\vbox to8pt{\hbox to0pt{\hss\box\capbox}\vss}}}
\def\cappar{\afterassignment\docappar\noexpand\let\a }

\begin{document}

\newcommand{\ee}{{\rm e}}
\newcommand{\dd}{{\rm d}}
\newcommand{\p}{\partial}
\newcommand{\calT}{{\cal T}}

\newcommand{\Jfree}{J^{\rm F}}
\newcommand{\Jfreeone}{J^{\rm free}_1}
\newcommand{\Jfreetwo}{J^{\rm free}_2}
\newcommand{\Jjam}{J^{\rm J}}
\newcommand{\Jjamone}{J^{\rm jam}_1}
\newcommand{\Jjamtwo}{J^{\rm jam}_2}
\newcommand{\Jf }{J^{\rm F}}
\newcommand{\Jj }{J^{\rm J}}
\newcommand{\Jff}{J^{\rm FF}}
\newcommand{\Jfj}{J^{\rm FJ}}
\newcommand{\Jjf}{J^{\rm JF}}
\newcommand{\Jjj}{J^{\rm JJ}}

\newcommand{\rhof }{\rho^{\rm F}}
\newcommand{\rhoj }{\rho^{\rm J}}
\newcommand{\rhoff}{\rho^{\rm FF}}
\newcommand{\rhofj}{\rho^{\rm FJ}}
\newcommand{\rhojf}{\rho^{\rm JF}}
\newcommand{\rhojj}{\rho^{\rm JJ}}
\newcommand{\zipper}{pairing\ }
\newcommand{\Zipper}{Pairing\ }

\newcommand{\rhofree}{\rho^{\rm free}}
\newcommand{\rhofreeone}{\rho^{\rm free}_1}
\newcommand{\rhofreetwo}{\rho^{\rm free}_2}
\newcommand{\rhojam}{\rho^{\rm jam}}
\newcommand{\rhojamone}{\rho^{\rm jam}_1}
\newcommand{\rhojamtwo}{\rho^{\rm jam}_2}

\newcommand{\alphaone}{\alpha_1}
\newcommand{\alphatwo}{\alpha_2}
\newcommand{\alphac}{\alpha_{\rm c}}

\newcommand{\la}{\langle}
\newcommand{\ra}{\rangle}
\newcommand{\beq}{\begin{equation}}
\newcommand{\eeq}{\end{equation}}
\newcommand{\bea}{\begin{eqnarray}}
\newcommand{\eea}{\end{eqnarray}}
\def\lsim{\:\raisebox{-0.5ex}{$\stackrel{\textstyle<}{\sim}$}\:}
\def\gsim{\:\raisebox{-0.5ex}{$\stackrel{\textstyle>}{\sim}$}\:}

\numberwithin{equation}{section}

\thispagestyle{empty}
\title{\Large {\bf 
Intersection of two TASEP traffic lanes\\[2mm]
with frozen shuffle update\\[2mm]
\phantom{XXX}
}}

\author{{C. Appert-Rolland, J. Cividini, and H.J.~Hilhorst}\\[5mm]
{\small 1 - University Paris-Sud; Laboratory of Theoretical Physics}\\
{\small B\^atiment 210, F-91405 ORSAY Cedex, France}\\[2mm]
{\small 2 - CNRS; UMR 8627; LPT}\\
{\small B\^atiment 210, F-91405 ORSAY Cedex, France}\\}

\maketitle
\begin{small}
\begin{abstract}
\noindent
Motivated by interest in pedestrian traffic 
we study two lanes (one-dimensional lattices) of length $L$
that intersect at a single site.
Each lane is modeled by a TASEP (Totally Asymmetric Exclusion Process).
The particles enter and leave
lane $\sigma$ (where $\sigma=1,2$) with probabilities
$\alpha_\sigma$ and $\beta_\sigma$, respectively.
We employ the `frozen shuffle' update introduced
in earlier work [C. Appert-Rolland et al, J. Stat. Mech. (2011) P07009],
in which
the particle positions are updated in a fixed random order.
We find analytically that each lane may be in a `free flow' 
or in a `jammed'
state. Hence the phase diagram in the domain
$0\leq\alpha_1,\alpha_2\leq 1$ 
consists of four regions 
with boundaries depending on $\beta_1$ and $\beta_2$.
The regions meet in a single point 
on the diagonal of the domain.
Our analytical predictions for the phase boundaries
as well as for the currents and densities in each phase
are confirmed by Monte Carlo simulations.
\\

\end{abstract}
\end{small}
\vspace{12mm}

\noindent LPT Orsay 11/xx
\newpage


\section{Introduction} 
\label{sect_introduction}

\cappar
Pedestrian motion has raised increasing interest
in the past years, both from a practical and a theoretical
point of view.
Understanding the behavior of crowds or of waiting lines
is still a challenge.
Simplified models may help to understand the behavior
of individuals and the resulting collective behavior
in various settings.
In a large class of models
\cite{schadschneider_k_n03,burstedde01b}
pedestrians are represented as hard core 
particles moving on a lattice according to
certain rules of motion.
One important ingredient of these rules
is the type of update scheme that is employed.
Actually the update scheme is an integral part of the model
definition;
changing the scheme may change the interpretation and the
properties of the model \cite{rajewski98}. 

In the past two types of updates have been used for
pedestrians modeling: the random shuffle update
\cite{wolki_s_s06,wolki_s_s07,smith_w07a,klupfel07}
which has been later replaced by the parallel update
\cite{kirchner_n_s03,kirchner03}.
In \cite{appert-rolland_c_h11a,appert-rolland_c_h11b}
we have proposed a new update scheme for pedestrian
modeling, the frozen shuffle update, that we shall
use in this paper.
Its characteristic feature 
is that during each time step all particles present in the system are
updated in a fixed random sequence. Newly entering particles are inserted
in this updating sequence and exiting particles are deleted from it 
according to a suitable algorithm.
Frozen shuffle update was inspired originally by the need
for a physically motivated rule of priority in cases where more than one
particle attempts to hop simultaneously towards the same target site. 
This update has the additional advantages that
it is easily implemented in a Monte Carlo simulation
and lends itself well to analytic study.
\vspace{2mm}

The consequences of frozen shuffle update
were worked out previously for the case of a one-dimensional
totally asymmetric exclusion process (TASEP)
both on a ring \cite{appert-rolland_c_h11a} and with open 
boundary conditions \cite{appert-rolland_c_h11b}. 
On a finite one-dimensional lattice with open boundaries
two parameters $\alpha$ and $\beta$ describe
the probabilities for particles to enter the system at one end and to
leave it at the other end. In this case the particle density $\rho$ and
the current $J$ must be determined as a function of $\alpha$ and
$\beta$. For varying $\alpha$ there appears to be
a critical point $\alpha=\beta$ between a `free flow' and a 
`jammed' state.
\vspace{2mm}

One of our purposes is to model pedestrian motion 
at the intersection of two corridors or two streets and to study
how global structures emerge from local interactions.
As a step toward this goal we
study in the present paper a TASEP on two perpendicular traffic lanes, 1 and 2,
that intersect at a single lattice site and whose entrance and exit
parameters are $\alpha_1, \alpha_2, \beta_1,$ and $\beta_2$.
The main question is again to determine 
the stationary state currents ${J}_1$ and ${J}_2$ in 
this two lane system as a function of these four parameters. 
For each lane one may expect two possibilities, 
a free flow (F) or a jammed (J) state.
We will study the phase diagram in the $\alpha_1\alpha_2$ plane
for $0\leq\alpha_1,\alpha_2\leq 1$,
considering $\beta_1$ and $\beta_2$ as fixed parameters.
Indeed we find a division of 
this square domain into four
different regions denoted FF, FJ, JF, and JJ, and separated by phase
boundaries for which we obtain analytic expressions.
\vspace{2mm} 

In earlier analytic work \cite{appert-rolland_c_h11a}
on the frozen shuffle update 
the concept of a `platoon' was introduced%
\footnote{This term has been borrowed from road traffic.}.
It will again play a role in this paper.
We will, moreover, point out here a new phenomenon, 
to be called the `pairing mechanism', which is operative at the intersection.
It says, basically, that when both lanes are in the jammed
state,
a platoon crossing the intersection on lane $1$ is always
accompanied by a platoon crossing the intersection on lane $2$.
This mechanism, 
which is an unintended consequence of the rules of motion,
will enable us to extend the theoretical
analysis from the single lane to the case of two intersecting lanes.
\vspace{2mm}

This article is organized as follows.
In section \ref{sect_rulesofmotion} we define 
the exact rules of motion of
this intersecting lane model and recall the concept of a `platoon'.
In a short section \ref{sect_onelane} we recall the single lane
results that will be needed again here.
In subsection \ref{sect_ffphaseintro} we argue that the FF phase, expected
to exist at low entrance rates, cannot extend beyond a certain
curve in the $\alpha_1\alpha_2$ plane.
In subsection \ref{sect_zipper} we show that if a JJ phase exists,
the exiting flow must obey a \zipper mechanism.
Exploiting this mechanism we determine in
subsections \ref{sect_jjphase} and \ref{sect_ffphase} 
the phase boundaries of the intermediate FJ/JF states
with the JJ and FF states, respectively, and thereby confirm the
existence of all four possible phases.  We obtain analytical
expressions for the currents in both lanes in each of
the four phases.
In subsection \ref{sect_limits} 
we consider various limits in the $\alpha_1\alpha_2$ domain.
In section \ref{sect_density} we derive expressions for the particle
densities which, in contrast to the currents,
are discontinuous at the phase boundaries.
In section \ref{sect_simulations} we present a few simulation results.
The data for the current fall right onto the theoretical
curves,
whereas the density data show finite size effects
similar to those encountered and explained in the single lane case
\cite{appert-rolland_c_h11b}.
Section \ref{sect_conclusion} is our conclusion.

\begin{figure}
\begin{center}
\scalebox{.45}
{\includegraphics{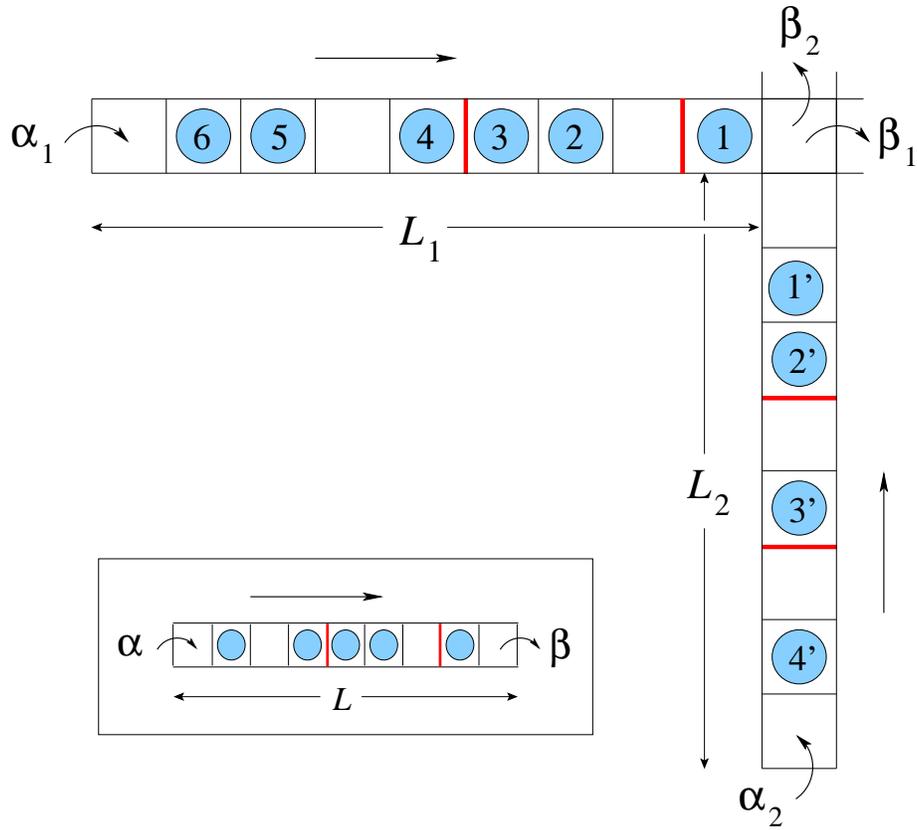}}
\end{center}
\caption{\small The two lane configuration studied in this work.
The arrows indicate the flow direction.
A heavy (red) bar indicates an end-of-platoon.
The entrance probabilities are $\alpha_1$ and $\alpha_2$,
the exit probabilities from the intersection site are $\beta_1$ and
$\beta_2$.\,\, 
Inset: the single lane with parameters $\alpha$ and $\beta$
studied in reference \cite{appert-rolland_c_h11b}.
}
\label{fig_twocross}
\end{figure}


\section{Rules of the motion}
\label{sect_rulesofmotion}


\subsection{Rules}
\label{sect_rules}

We consider the  geometry shown in figure \ref{fig_twocross},
consisting of two perpendicular
one-dimensional lattices (or: lanes) labeled by an index $\sigma=1,2$.
Hard core particles may move to the right on lane 1 and upward on lane 2.
The lanes have $L_1$ and $L_2$ sites, respectively, plus
a common intersection site.
Particles enter at the two extremal sites and exit
when leaving the intersection site.
In our analytical treatment we shall 
imagine that both $L_\sigma$ tend to infinity.
As usual in critical phenomena,
finite size effects will be important
only in the critical region.
In practice, our infinite system results apply
as soon as the $L_\sigma$ are large compared to the boundary
layers near the entrance or exit.

A particle $i$, when entering the system, is assigned,
in a way discussed below, a phase%
\footnote{We use in this work the term `phase' both to designate
  the $\tau_i$ assigned to the particles
  and to refer to the different types of stationary
  states of the system as a whole. No confusion need arise.}
$\tau_i\in[0,1)$ 
which it keeps as a fixed attribute until it exits the system.
Although the time $t$ is continuous,
the time evolution is best described in terms of integer
time steps $s=1,2,3,\ldots$.
During each time step the particles are visited
in the order of increasing phases%
\footnote{For a closed system with a fixed number $N$ of particles 
a randomly chosen permutation of the particles
may replace the assignment of phases \cite{appert-rolland_c_h11a}.}
and their positions are updated according to the following rules.

{\it General rule.}\,\,
When a lane 1 (lane 2) particle is updated, it moves one lattice
distance to the right (upward) if the target site is
empty\footnote{The TASEP considered in this paper is
{\em deterministic} in the bulk.
For particles hopping forward to an empty target site with a probability
$p<1$, analytical predictions would probably be more difficult.},
and does not move if the target site is occupied. 
The update of particle $i$ during the
$s$th time step is considered to
take place at the exact moment $t=s+\tau_i$\,. 

The general rule must be supplemented by two special rules for
entering and exiting particles.

{\it Exiting rule.}\,\, 
A particle $i$ which at the beginning of the $s$th time step is on the 
intersection site, 
will at time $s+\tau_i$ leave the system with probability 
$\beta_1$ (or $\beta_2$)
according to whether it has arrived through lane 1 (or lane 2).
Once it leaves the intersection site, we do not consider it any longer:
it has left the system%
\footnote{In fact, it has only left our window of observation:
we could consider that each lane extends 
beyond the point of intersection, and that the particle,
once it leaves the intersection, 
enters a free flow state where
it continues to move at every time step.}.

{\it Entering rule.}\,\,
When the entrance site of lane $\sigma$ becomes vacant 
at time $t$, it will be occupied
by a new particle, injected from outside, at a random time $t'=t+T$. 
Here $T$ is drawn from the exponential distribution
\beq
P_\sigma(T) = a_\sigma\,\ee^{-a_\sigma T}, \qquad 0\leq T < \infty,
\label{dPT}
\eeq
where the `entrance rate' $a_\sigma>0$ is a model parameter.

Equivalent to $a_\sigma$ is the `entrance probability' $\alpha_\sigma$
defined by
\beq
\alpha_\sigma = 1-\ee^{-a_\sigma}, \qquad \sigma=1,2,
\label{da2}
\eeq
which is the conditional probability that the entrance site of lane $\sigma$
is occupied at time $t+1$ given that it was vacant at time $t$.
Henceforth we will sometimes
let $a_\sigma$ and 
$\alpha_\sigma$ occur in the same expression.

To see the motivation for the above entering rule, one may notice
that each particle arrival on the entrance site (or, for that matter,
on any other site) is followed by one unit of `dead time' 
during which no new arrival on that site is possible.
Subject to this dead time condition,
the entering rule distributes
the instants of arrival of the particles at the entrance site
uniformly on the time axis
\footnote{An analogy with a one-dimensional system of hard rods of
unit length was pointed out in reference \cite{appert-rolland_c_h11b}.}.


\subsection{Platoons}
\label{sect_platoons}

The phases $\tau_i$ may be regarded as quenched random variables.
With the rules stated above the particle motion is deterministic apart from 
the stochastic phase assignment at the entrances and,
when $\beta_1<1$ or $\beta_2<1$, the stochastic exits. 
As a consequence of the entering rule
the phase $\tau'$ of the newly injected particle is
related to the phase $\tau$ of its predecessor in the same lane by
\beq
\tau'=(\tau+T)\,{\rm mod}\,1.
\label{reltauprimetau}
\eeq 
It follows that there are correlations%
\footnote{Described in detail in reference \cite{appert-rolland_c_h11b}.}
between the phases of successive particles in the same lane.
Following a lane 
in the direction opposite to the particle flow,
one may group the particles together into sequences of
increasing phases, a phase decrease signaling the beginning of a
new sequence.
When particles constituting such an increasing phase sequence occupy
consecutive sites, they are said to constitute a `platoon'.
The average platoon length $\nu$ associated with an
entrance probability $\alpha$
is given by \cite{appert-rolland_c_h11b}
\beq
\frac{1}{\nu(\alpha)} = 1 + \frac{1}{a} - \frac{1}{\alpha}\,. 
\label{xnua}
\eeq
where $\alpha\equiv 1-\ee^{-a}$.
This quantity will play an essential role in the analysis that follows.


\section{Stationary states in a single lane}
\label{sect_onelane}

The single lane problem 
with boundary conditions $\alpha$ and $\beta$, shown in
the inset of figure \ref{fig_twocross},
has yielded \cite{appert-rolland_c_h11b}
results some of which will again be needed here.
First, we know 
that the entrance probability $\alpha$ (for large enough $\beta$)
imposes a `free flow' bulk state%
\footnote{Except for a jammed boundary layer of fluctuating size
  near the exit.}, 
that is, one in which all attempted moves are successful, 
which carries a current
\beq
\Jf(\alpha)=\frac{a}{1+a}\,.
\label{xJfree1}
\eeq
Secondly,
it was shown \cite{appert-rolland_c_h11b} that 
the exit probability $\beta$ (for large enough $\alpha$)
imposes a jammed bulk state%
\footnote{Except for a free flow boundary layer of fluctuating size
  near the entrance.  
  Only in the limit $L\to\infty$ are the two phases
  well-defined in the sense that tunneling between them becomes
  impossible.}.
This is a state in which all particles belong to platoons
and successive platoons are separated by at most a single vacancy.
The jammed state has an outgoing current 
\beq
\Jj(\alpha,\beta) = \frac{\nu}{ \frac{\nu}{\beta}+1 }\,, 
\label{xJjam1}
\eeq
with $\nu$ determined by $\alpha$ through (\ref{xnua}).
This exit flow can be sustained if and only if the entering flow
is sufficiently large, that is for $\Jf>\Jj$. 
The equality $\Jf(\alpha)=\Jj(\alpha,\beta)$ 
therefore defines a critical point, which turns out to occur for
$\alpha=\beta$. As a consequence, the stationary state current $J$ is 
equal to $J=\Jf(\alpha)$ for $\alpha\leq\beta$ and $J=\Jj(\alpha,\beta)$ for
$\alpha\geq\beta$; at the critical point it is continuous but undergoes
a change of slope.
Finally, for $\alpha=\beta$ the two phases coexist in the
system and are spatially separated by a sharp domain wall.


\section{Phase diagram of the two lane system}
\label{sect_twolanes}


\subsection{The FF phase}
\label{sect_ffphaseintro}

In the two lane system 
the entrance probabilities $\alpha_1$ and
$\alpha_2$ strive to impose independent free flow states in each lane,
that is, an FF phase with currents%
\footnote{A symbol with a single upper index, F or J, refers to
  an auxiliary one lane system; a symbol with a double upper index refers to 
  one of the lanes $\sigma=1,2$ of the two lane system under study.}
\bea
\Jff_\sigma &=& \Jf(\alpha_\sigma) \nonumber\\[2mm]
            &=& \frac{a_\sigma}{1+a_\sigma}, \qquad \sigma=1,2.
\label{xJfree2}
\eea
The two currents interact at the intersection site
where moreover they are subject to random exits with probabilities $\beta_1$
and $\beta_2$.
We anticipate that if at given  $\beta_1$ and $\beta_2$ 
the entrance probabilities $\alpha_1$ and $\alpha_2$ become small enough, 
the system will be in an FF phase.
The interaction between the currents at
the exit site may then occasionally
delay individual particles, but will not create 
waiting queues that grow without limit.

However, the rules of the motion are such that
at each time step at most a single particle can leave the exit site.
This immediately yields a bound 
for the FF phase  in the $\alpha_1\alpha_2$ plane:
whenever $\Jf(\alpha_1)+\Jf(\alpha_2)>1$, 
there must necessarily occur 
formation of an ever growing waiting line in at least one of the two
lanes and the system cannot then be in its FF phase.
This condition may be rewritten as $a_1a_2>1$.
In subsection \ref{sect_ffphase} we will show by explicit
calculation that the FF phase does indeed
exist and analytically determine its phase boundary.


\subsection{Pairing mechanism}
\label{sect_zipper}

There is no standard way of finding the
phase diagram of this two lane system. 
We therefore develop following reasoning.

\begin{figure}
\begin{center}
\scalebox{.45}
{\includegraphics{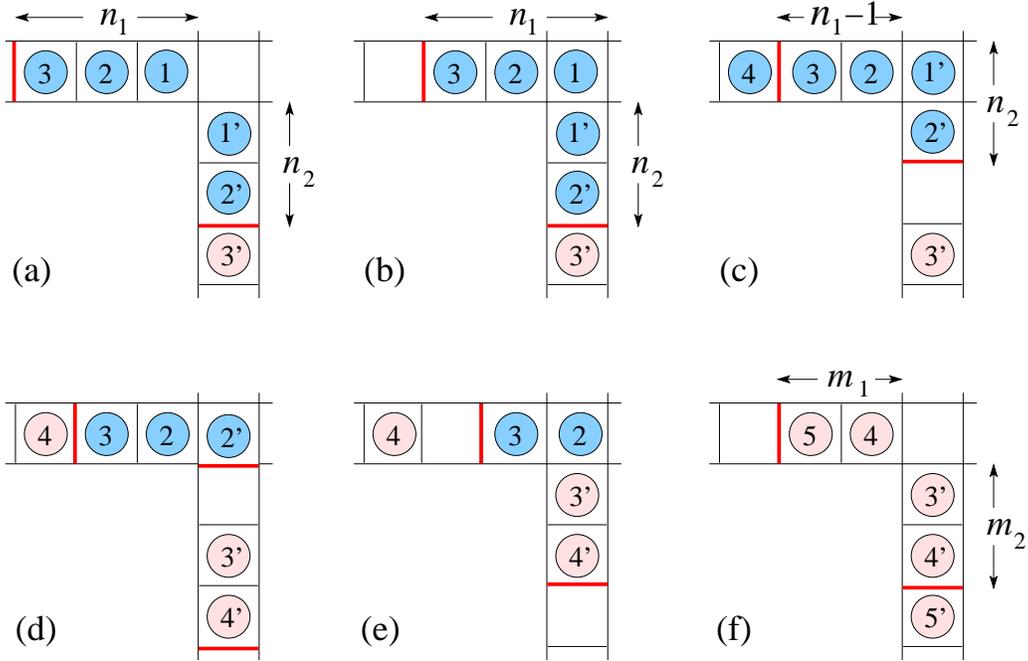}}
\end{center}
\caption{\small Sequence of snapshots illustrating the pairing effect
of platoons at the intersection site. 
In this example $n_1=3$ and $n_2=m_1=m_2=2$. See text.
}
\label{fig_zipper}
\end{figure}

Let us suppose now that both lanes are jammed, that is, the system is
in a JJ state.
In the two-lane problem there then appears
a new phenomenon. 
The rules of the motion have an unintended consequence that we will
call the {\it pairing mechanism}. 
This is the phenomenon that the exiting platoons of the two lanes are
rigorously paired:
for each platoon exiting from lane 1 there
is also a platoon exiting from lane 2, and {\it vice versa}.
To demonstrate this effect we refer to figure \ref{fig_zipper}.

Figure \ref{fig_zipper}a shows a particle configuration near the exit
at some integer instant of time $t=s$ at which the
intersection site is empty.
A heavy (red) bar (below or to the left of a particle) 
marks an end-of-platoon. 
In each lane there is a platoon (dark colored particles) waiting to
enter the empty intersection site;
the platoons have lengths $n_1$ and $n_2$ 
and are headed by the particles marked $1$ and $1^\prime$.
At the next time step, $t=s+1$, the platoon head with the lower phase will 
hop onto the intersection site, its whole platoon will advance by
one lattice distance, and it will block the other waiting particle.
Let us suppose, as shown in figure \ref{fig_zipper}b, 
that it is the horizontally moving platoon that advances.
During each of the next time steps, 
the lane 1 particle (marked 1), as long as it occupies the exit site,  
will exit with probability $\beta_1$, so it 
will exit only after on average $\beta_1^{-1}$ time steps.
During the same time step in which it exits, 
out of the two particles (marked $2$ and $1^\prime$) 
that are waiting to hop onto
the exit site, again the one with the lower phase will 
effectively hop, 
pull along its whole platoon, 
and block the other one. 
Let us suppose that it is the lane 2 particle (marked 1$^\prime$)
that advances. This takes us to the configuration of 
figure \ref{fig_zipper}c. It is equivalent to the one of 
\ref{fig_zipper}b,
except for an interchange of the roles of lanes 1 and 2,
accompanied by the replacements $n_1 \mapsto n_2$ and 
$n_2 \mapsto n_1-1$. The procedure taking us form \ref{fig_zipper}b to 
\ref{fig_zipper}c will now repeat itself
{\it mutatis mutandis\,} and each time either $n_1$ or $n_2$ will
decrease by one unit.
Lane 1 and lane 2 particles have average exit times  
$\beta_1^{-1}$ and $\beta_2^{-1}$, respectively.
At some point
the last particle of one of the two platoons is on the intersection site. 
Let us suppose this is a lane 2 particle, as in \ref{fig_zipper}d
where it is marked $2^\prime$. 
During the same the time step in which $2^\prime$  leaves the
intersection site,
that site will be occupied by the particle waiting in the other
lane (marked $2$), which has a higher phase. 
This will result in the situation of figure \ref{fig_zipper}e.
The next lane 2 particle (marked $3^\prime$) belongs to the
next platoon; if it has
already arrived at the waiting site
(which may or may not be the case), it will be blocked in that time
step. It will similarly be blocked in all following time steps,
until the last remaining particle (marked $3$)
of the lane 1 platoon leaves the intersection site. 
The situation that then results is depicted in figure \ref{fig_zipper}f.
It is identical to that of figure \ref{fig_zipper}a, except that now
the next two platoons, 
of lengths $m_1$ and $m_2$ and headed by particles $4$ 
and $3^\prime$, are waiting to enter the intersection site.
This is the pairing effect.

We remark parenthetically that
this pairing argument is easily extended to an arbitrary number $p$ of
lanes intersecting at a single site, when they are all in the jammed phase.
We will not, however, attempt to consider here such more general geometries.
\vspace{2mm}

We will now exploit this effect to find an expression for the current
in the JJ phase.
In order to arrive at the situation of figure \ref{fig_zipper}f
starting from the one of figure \ref{fig_zipper}a there 
is first the time step in which the exit site gets occupied.
Next, there are $n_1$ lane 1 particles and $n_2$ lane 2 particles that
leave the intersection site subject to the exit probabilities $\beta_1$ and
$\beta_2$, respectively. The total time $t_{n_1n_2}$
needed for this process and averaged over all exit histories therefore 
is $t_{n_1n_2}= n_1\beta_1^{-1} + n_2\beta_2^{-1} + 1$.
Let $\nu_1\equiv\nu(\alpha_1)$ and $\nu_2\equiv\nu(\alpha_2)$ 
be the average platoon lengths in the two lanes. 
Then the mean exit time $t_{\rm exit}$ of an arbitrary 
pair of platoons to exit is the
average of $t_{n_1n_2}$ over all platoon lengths. This yields
\beq
t_{\rm exit} = \frac{\nu_1}{\beta_1}+\frac{\nu_2}{\beta_2}+1.
\label{exittime}
\eeq
Hence in the JJ phase the outgoing currents $\Jjj_1$ and $\Jjj_2$ 
of the two lanes are given by
\beq
\Jjj_\sigma = \frac{\nu_\sigma}{ \frac{\nu_1}{\beta_1} + 
\frac{\nu_2}{\beta_2} + 1 }, 
\qquad \sigma=1,2.
\label{xJjam2}
\eeq
This expression is a nontrivial generalization of the single lane formula 
(\ref{xJjam1}).
Both currents (\ref{xJjam2}) depend on all four parameters 
$\alpha_1$, $\alpha_2$, $\beta_1$, $\beta_2$.
The ratios $\nu_\sigma / \beta_\sigma$, which will reappear frequently
below, show the scaling with $\beta_\sigma$ of the time that the
intersection site is occupied by lane $\sigma$ particles.

All elements are in place now for us to go on and find the phase
boundaries in the domain $0\leq\alpha_1,\alpha_2\leq 1$.


\subsection{Boundaries of the JJ and FJ/JF phases}
\label{sect_jjphase}

\begin{figure}
\begin{center}
\scalebox{.55}
{\includegraphics{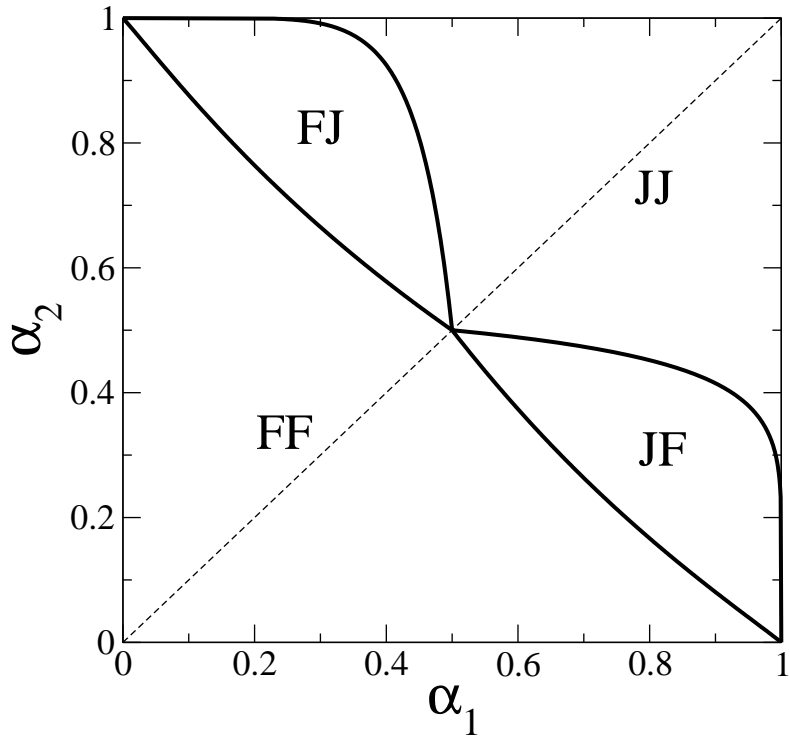}}
\end{center}
\caption{\small 
Phase diagram in the $\alpha_1\alpha_2$ plane for $\beta_1=\beta_2=1$.
The heavy solid lines are phase boundaries.
Dashed line: the diagonal. 
The symbols F (free) and J (jammed) refer to lanes 1 and 2
in the order given. The four-phase point is at 
$(\alpha_1,\alpha_2)=(\tfrac{1}{2},\tfrac{1}{2})$.
}
\label{fig_phdiagsymm}
\end{figure}

\begin{figure}
\begin{center}
\scalebox{.55}
{\includegraphics{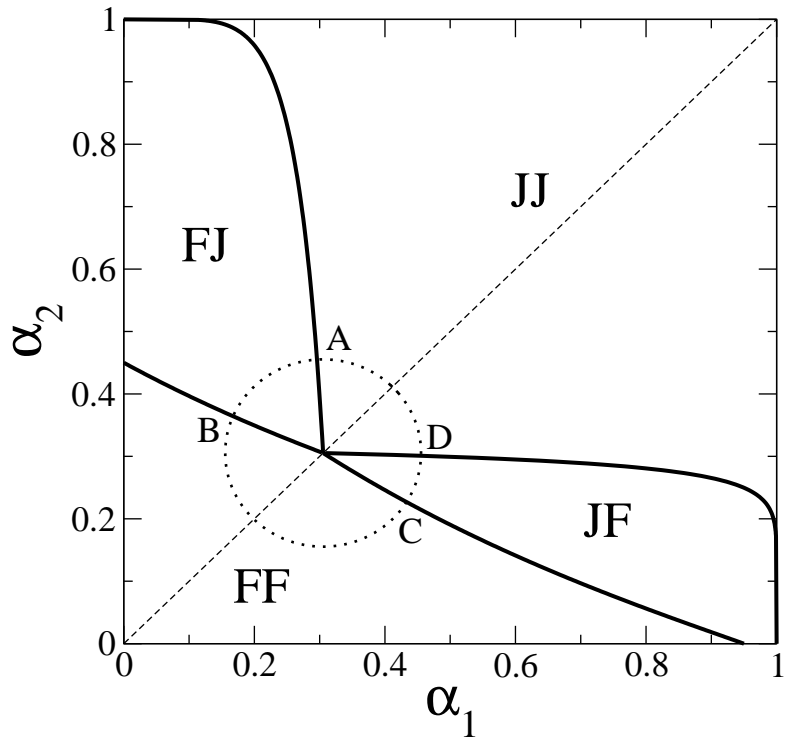}}
\end{center}
\caption{\small 
Phase diagram in the $\alpha_1\alpha_2$ plane for $\beta_1=0.95$ and
$\beta_2=0.45$. All symbols are as in figure \ref{fig_phdiagsymm}.
The four-phase point is on the diagonal at $(\alphac,\alphac)$
with $\alphac$ given by (\ref{xalphac}).
As shown in section \ref{sect_boundariesFF},
the boundary curves of the FF phase intersect the $\alpha_1$ and
$\alpha_2$ axes in $\alpha_1=\beta_1$ and $\alpha_2=\beta_2$, respectively.
Dotted circle: path along which the simulation
results of figures \ref{fig_circleplotJ} 
and \ref{fig_circleplotrho} were obtained.
}
\label{fig_phdiagasymm}
\end{figure}

Let us suppose the system is in the JJ phase.
The condition for the system to be able to sustain these 
queues is that in both lanes the
out-current $\Jjj_\sigma$
be smaller than the corresponding free flow 
entrance driven current
$\Jf(\alpha_\sigma)$. 
That is, for the JJ phase to be stable we should satisfy the two inequalities
\beq
\Jf(\alpha_\sigma) \geq \Jjj_\sigma\,, \qquad \sigma=1,2,
\label{conditionsjj0}
\eeq
or, upon substituting (\ref{xJfree2}) and (\ref{xJjam2}) 
in (\ref{conditionsjj0}),
\beq
\frac{a_\sigma}{1+a_\sigma} \geq 
\frac{\nu_\sigma}{ \frac{\nu_1}{\beta_1} + \frac{\nu_2}{\beta_2} + 1 }\,,
\qquad \sigma=1,2.
\label{conditionsjj1}
\eeq
On the borderline of the JJ phase equation (\ref{conditionsjj1}) should hold
as an equality for either $\sigma=1$ or $\sigma=2$.
To rewrite this equality we invert both of its members
and use (\ref{xnua}).
It then follows that
\beq
\frac{\nu_\sigma}{\alpha_\sigma} = \frac{\nu_1}{\beta_1} +
\frac{\nu_2}{\beta_2}\,,
\qquad \sigma=1,2,
\label{conditionsjj2}
\eeq
or, equivalently, 
\begin{subequations}\label{conditionsjj3}
\bea
\nu_1\left( \frac{1}{\alpha_1}-\frac{1}{\beta_1} \right) &=& 
\frac{\nu_2}{\beta_2}\,, 
\label{condjj3a} \\[2mm]
\nu_2\left( \frac{1}{\alpha_2}-\frac{1}{\beta_2} \right) &=& 
\frac{\nu_1}{\beta_1}\,. 
\label{condjj3b} 
\eea
\end{subequations}
Equations (\ref{conditionsjj3}) 
represent two intersecting curves in the
$\alpha_1\alpha_2$ plane. Although they depend on the
parameters $\beta_1$ and $\beta_2$, their point of intersection
always lies on the diagonal $\alpha_1=\alpha_2$.
To see this, note that $\nu_\sigma=\nu(\alpha_\sigma)$
is a function only of $\alpha_\sigma$ and hence
$\alpha_1=\alpha_2=\alphac$ implies that $\nu_1=\nu_2=\nu_{\rm c}$.  
Using this in (\ref{conditionsjj3}) we see that $\nu_{\rm c}$ divides out
and that both equations are satisfied by
\beq
\alphac = \frac{\beta_1\beta_2}{\beta_1+\beta_2}\,.
\label{xalphac}
\eeq
Hence equation (\ref{condjj3a}) gives the JJ/FJ boundary 
in the triangle above the diagonal $\alpha_1=\alpha_2$ 
and (\ref{condjj3b}) gives the JJ/JF boundary 
in the triangle below this diagonal.
For the special case $\beta_1=\beta_2=1$ these phase boundaries
are shown in figure \ref{fig_phdiagsymm}, which is symmetric with respect to
the diagonal.
An example of the general case with $\beta_1\neq\beta_2$ 
is shown in figure \ref{fig_phdiagasymm}, where this symmetry has been lost.

\subsection{Boundaries of the FF  and FJ/JF phases}
\label{sect_ffphase}

We wish to find now the borderline between the FJ and JF phases, on
the one hand, and the FF phase on the other hand. 
To be definite, let us suppose the system is in the FJ phase
so that we know that the current in lane 1 is given by its free flow
expression 
\beq
\Jfj_1=\Jfree(\alpha_1).
\label{xJfj1}
\eeq 
The current $\Jfj_2$ in lane 2 is, however, unknown. 
In order to calculate $\Jfj_2$ we cannot invoke 
now the pairing mechanism, since it requires {\it both\,} lanes to be
jammed. Instead, let us suppose that for every platoon that exits the jammed 
lane 2, and
that is known to contain on average $\nu_2$ walkers, there are on 
average $\mu_1$ walkers that exit lane 1. Here $\mu_1$ is unknown but 
necessarily satisfies\footnote{Relation (\ref{relmunu}), 
valid in the FJ phase, becomes an equality when lane 1 also gets jammed,
so that $\mu_1=\nu_1$ should give again the JJ/FJ boundary.
This may be verified by explicit calculation.}

\beq
\mu_1 \leq \nu_1\,.
\label{relmunu}
\eeq
Then we have for the exiting currents in the FJ phase
the two asymmetric expressions
\begin{subequations}\label{xJjamfj}
\bea
\Jfj_1 &=& \frac{\mu_1}{ \frac{\mu_1}{\beta_1}+\frac{\nu_2}{\beta_2}+1 }\,,
\label{xJjamfj1}
\\[2mm]
\Jfj_2 &=& \frac{\nu_2}{ \frac{\mu_1}{\beta_1}+\frac{\nu_2}{\beta_2}+1 }\,.
\label{xJjamfj2}
\eea
\end{subequations}
Upon combining (\ref{xJfj1}) with (\ref{xJjamfj1}) and using
(\ref{xJfree1}) we may solve for $\mu_1$ and find
\beq
\mu_1 = \frac{ \left( \frac{\nu_2}{\beta_2}+1 \right)a_1 }
{1 + \left( 1-\frac{1}{\beta_1} \right)a_1}\,.
\label{xmu1}
\eeq
The condition for the sustainability of the jammed phase in lane 2 is
\beq
\Jfree(\alpha_2) \geq \Jfj_2\,.
\label{conditionsfj0}
\eeq
The phase transition line is obtained when (\ref{conditionsfj0}) 
holds as an equality,
which, with the substitutions of (\ref{xJjamfj2}) and (\ref{xJfree1}), 
happens for
\bea
\frac{a_2}{1+a_2} &=& \frac{\nu_2}
{\frac{\mu_1}{\beta_1}+\frac{\nu_2}{\beta_2}+1}
\nonumber\\[2mm]
&=& \nu_2\,
\frac{a_1}{1+a_1}\,\frac{1+\left( 1-\frac{1}{\beta_1} \right)a_1}
{\left( \frac{\nu_2}{\beta_2}+1 \right)a_1}
\nonumber\\[2mm]
&=& \frac{\nu_2\beta_2}{\nu_2+\beta_2}\,
\frac{1+\left( 1-\frac{1}{\beta_1} \right)a_1}{1+a_1}\,,
\label{conditionsfj1}
\eea
where to pass from the first to the second line we first noticed that by
(\ref{xJjamfj1}) and (\ref{xJfj1}) the denominator on the RHS is equal to
$\mu_1/\Jfj_1=\mu_1/\Jf(\alpha_1)=\mu_1(1+a_1)/a_1$ 
and then substituted for $\mu_1$ 
expression (\ref{xmu1}).
We may solve (\ref{conditionsfj1}) for $a_1$ and find
\beq
a_1 = \frac{\beta_1(1-R_2)}{1-\beta_1(1-R_2)}\,,
\label{xa1}
\eeq
where we introduced the abbreviation $R_2\equiv R(\alpha_2,\beta_2)$ with
\beq
R(\alpha,\beta) = 
\frac{a}{1+a}\,\frac{\nu+\beta}{\nu\beta}\,.
\label{dR2}
\eeq
We will continue to consider $\beta_1$ and $\beta_2$ as fixed parameters.
When in (\ref{xa1}) 
we use (\ref{dR2}) for $R_2$,
(\ref{xnua}) for $\nu_2$, and (\ref{da2}) for $\alpha_2$,
it becomes an explicit solution for $a_1$ in terms of $a_2$,
or equivalently, for $\alpha_1$ in terms of $\alpha_2$\,.
Hence (\ref{xa1}) constitutes our final result for the  FF/FJ boundary.
A permutation of indices gives the FF/JF boundary.
These phase boundaries are again shown in figures \ref{fig_phdiagsymm} and
\ref{fig_phdiagasymm} for the symmetric case with $\beta_1=\beta_2=1$ 
and for a typical asymmetric case, respectively.


\subsection{Limiting cases}
\label{sect_limits}

We consider in this section the limit behavior of the phase boundaries
as they approach the borders of the domain $0\leq\alpha_1,\alpha_2\leq 1$.


\subsubsection{Boundaries of the JJ phase}
\label{sect_boundariesJJ}

One obtains from (\ref{condjj3a}) 
the behavior of the JJ/FJ boundary in the limit of small $\alpha_1$
by noticing that 
$\lim_{\alpha_1\to 0}\nu_1=2$ and that therefore in that limit the LHS of 
(\ref{condjj3a}) diverges, which forces $\nu_2$ on the RHS also to diverge.
Using next that for $\alpha_2\to 1$ one has
$\nu_2 \simeq -\log(1-\alpha_2)$ one finds
\beq
\alpha_2 \simeq 1-\ee^{-2\beta_2/\alpha_1}, \qquad \alpha_1\to 0.
\label{xalpha1to0}
\eeq
A permutation of indices gives the asymptotic
behavior of the JJ/JF boundary in the limit $\alpha_2\to 0$.
The exponentials of the inverse functions $1/\alpha_1$ and $1/\alpha_2$
explain the extremely rapid alignment of these curves
along the edges of the figure.


\subsubsection{Boundaries of the FF phase}
\label{sect_boundariesFF}

We wish to find the point of intersection of the FF/JF (FF/FJ) boundary
with the horizontal (vertical) axis. It is located
at $\alpha_1=\beta_1$ (at $\alpha_2=\beta_2$).
To show this, 
we ask what the limiting
value of $\alpha_2$ is when $\alpha_1\to 0$.
It is useful to notice that the quantity $R_2$ that occurs in
(\ref{xa1}) may be expressed as a ratio of two single lane currents,
\beq
R_2 = \frac{ J^{\rm free}(\alpha_2)}{J^{\rm jam}(\alpha_2,\beta_2)}\,.
\label{xR2}
\eeq
The `physical' argument goes as follows. For $\alpha_1=0$ lane 1
is unoccupied and the intersecting lane problem reduces to that of the
single open-ended lane with boundary conditions $(\alpha_2,\beta_2)$,
whose critical point is known \cite{appert-rolland_c_h11b}
to occur at $\alpha_2=\beta_2$.
Mathematically, $\alpha_1=0$ implies $a_1=0$; when this is substituted 
for the LHS of (\ref{xa1})
we find $R_2=1$, after which (\ref{dR2}) yields
$J^{\rm free}(\alpha_2)=J^{\rm jam}(\alpha_2,\beta_2)$.
When this equality is worked out 
we obtain the same result $\alpha_2=\beta_2$.
Finding the limit behavior for $\alpha_2\to 0$ amounts 
to a permutation of indices.

The straight line (not drawn in figures \ref{fig_phdiagsymm} and
\ref{fig_phdiagasymm}) that connects these two points of intersection
has the equation 
\beq
\frac{\alpha_1}{\beta_1} + \frac{\alpha_2}{\beta_2} = 1
\label{xstraight}
\eeq
and also passes through the critical point $(\alphac,\alphac)$.
In both figures, \ref{fig_phdiagsymm} and \ref{fig_phdiagasymm},
the boundary delimiting the FF phase is slightly curved and
falls just below this straight line.


\section{Particle density}
\label{sect_density}

The determination of the phase diagram was based exclusively on the
analysis of the particle currents in the different phases. 
From the preceding construction it follows that the currents are
continuous at the phase transition lines.
This differentiates them from the particle densities, which are
the quantities of interest in this section. 
We will denote a particle density generically by the symbol $\rho$, 
to which we attach indices according to the same convention as used
for $J$.
In all phases we have the relation $J=v\rho$, where $\rho$ is the
particle density and $v$ the average particle velocity. 
Since in the free flow phase all particles have velocity $v=1$,
we have $\rhof=\Jf$ and therefore
\beq
\rhoff_\sigma = \frac{a_\sigma}{1+a_\sigma}\,, \qquad \sigma=1,2,
\label{xrhoff}
\eeq
\beq
\rhofj_1 = \frac{a_1}{1+a_1}\,, \qquad \rhojf_2 = \frac{a_2}{1+a_2}\,.
\label{xrhof}
\eeq
In the jammed phase we have generically the
relation\footnote{This relation, derived in
\cite{appert-rolland_c_h11a}, is a direct consequence of the
structure of the jammed state
described in section \ref{sect_onelane}.} $J=\nu(1-\rho)$,
where $\nu$ is as before the average platoon length.
This gives the three relations
\beq
\Jjj_\sigma = \nu_\sigma(1-\rhojj_\sigma),
\qquad \sigma=1,2, 
\label{eqsrhojj}
\eeq
\beq
\Jfj_2 = \nu_2(1-\rhofj_2), \qquad \Jjf_1 = \nu_1(1-\rhojf_1).
\label{eqsrhoj}
\eeq
Solving these for the densities using the expressions found in section
\ref{sect_twolanes} for the currents we get
\beq
\rhojj_\sigma = \frac{ \frac{\nu_1}{\beta_1}+\frac{\nu_2}{\beta_2} }
{ \frac{\nu_1}{\beta_1}+\frac{\nu_2}{\beta_2}+1 }\,, 
\qquad \sigma=1,2,
\label{xrhojj}
\eeq
\beq
\rhofj_2 = \frac{ \frac{\mu_1}{\beta_1}+\frac{\nu_2}{\beta_2} }
{ \frac{\mu_1}{\beta_1}+\frac{\nu_2}{\beta_2}+1 }\,, 
\qquad
\rhojf_1 = \frac{ \frac{\nu_1}{\beta_1}+\frac{\mu_2}{\beta_2} }
{ \frac{\nu_1}{\beta_1}+\frac{\mu_2}{\beta_2}+1 }\,.
\label{xrhoj}
\eeq
Remarkably, equation (\ref{xrhojj}) shows that in the JJ phase the
particle densities in the two lanes are equal irrespective of the
values of $\alpha_1,\alpha_2,\beta_1,\beta_2$.

Equations (\ref{xrhoff}), (\ref{xrhof}), (\ref{xrhojj}), 
and (\ref{xrhoj})constitute our
analytic results for the particle densities in the four different phases.


\section{Simulations}
\label{sect_simulations}

\begin{figure}
\begin{center}
\scalebox{.55}
{\includegraphics{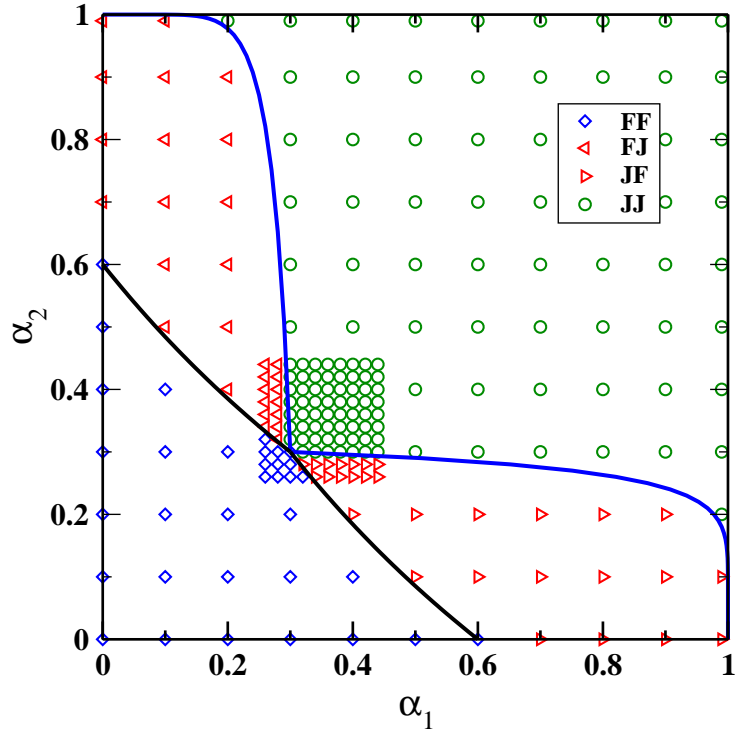}}
\end{center}
\caption{\small 
Phase diagram in the $\alpha_1\alpha_2$ plane for $\beta_1=\beta_2=0.6$.
The nature of the phase was determined (see text) on a grid of points
with a denser covering near the
four-phase point $(\alpha_1,\alpha_2)=(0.3,0.3)$.
The phase boundaries are those given by theory.
}
\label{fig_gridplot}
\end{figure}

In order to test the theory of the preceding sections we 
determined the phase of the system for fixed $\beta_1=\beta_2=0.6$
on a grid of points in the $\alpha_1\alpha_2$ plane. 
The grid was refined in a square region around the four-phase
point, which according to equation (\ref{xalphac})
occurs at $(\alpha_1,\alpha_2)=(0.3,0.3)$.
To determine the phase for a specific pair $(\alpha_1,\alpha_2)$, 
simulations were performed on intersecting lattices of lengths 
$L_1=L_2\equiv L=75$. In a finite system 
the entrance and exit boundary conditions try to impose different phases,
which as a consequence will be separated by a domain wall
\cite{appert-rolland_c_h11b,kolomeisky98,pigorsch_s00,santen_a02}.
Away from the critical point the fluctuating domain wall position
will be localized within some finite distance from one of the lane ends;
upon approach of criticality this localization length 
increases until it attains the lane length $L$.
In our simulations the
domain wall position was determined in each lane as described
in~\cite{appert-rolland_c_h11b}. We then
averaged it over 5\,000\,000 time steps after having first discarded
a transient of 5\,000 time steps in order to make sure that the system
was stationary.
The lane was classified F or J
if its mean position was closer to the exit or closer to the
entrance, respectively.
The results are represented in figure \ref{fig_gridplot}. They are in
perfect agreement with the theoretical phase boundaries, within the
resolution of the grid.

\begin{figure}
\begin{center}
\scalebox{.55}
{\includegraphics{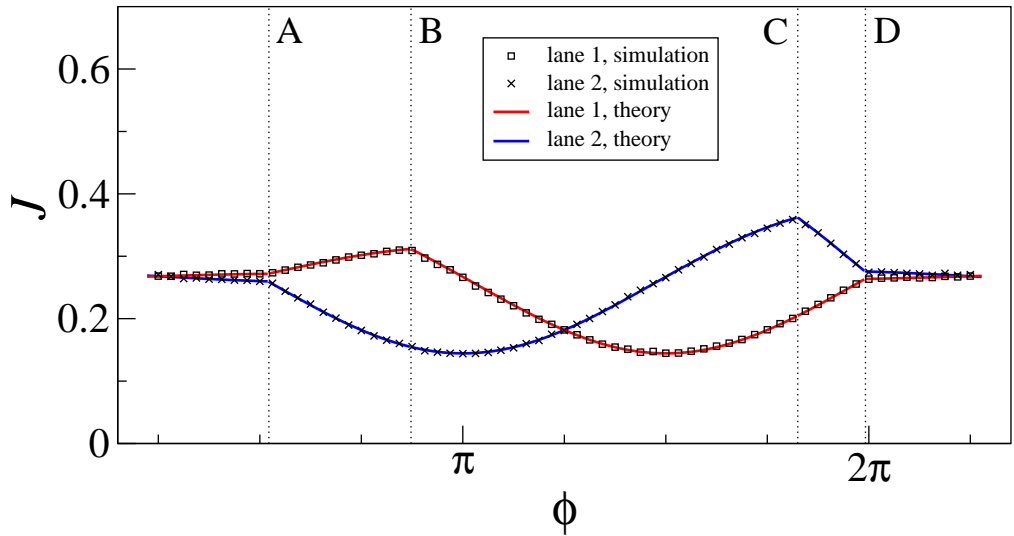}}
\end{center}
\caption{\small 
The current, denoted by the generic symbol $J$, 
as a function of the angle $\phi$ in lanes 1 and 2.
}
\label{fig_circleplotJ}
\end{figure}

\begin{figure}
\begin{center}
\scalebox{.55}
{\includegraphics{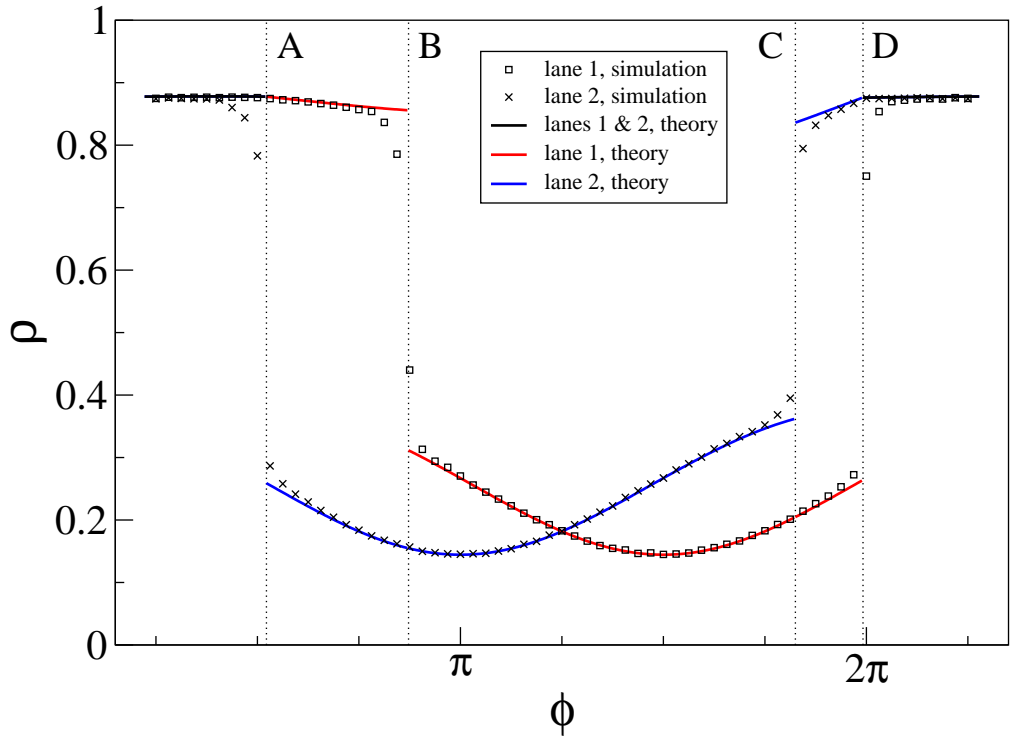}}
\end{center}
\caption{\small 
The density, denoted by the generic symbol $\rho$, 
as a function of the angle $\phi$ in lanes 1 and 2.
The black part of the theoretical curves is common to both lanes.
Comparison to figure \ref{fig_circleplotJ} 
shows that for a lane in the free flow phase we have $J=\rho$.
}
\label{fig_circleplotrho}
\end{figure}

A 
more detailed simulation was carried out for
the asymmetric case of figure \ref{fig_phdiagasymm}.
In the $\alpha_1\alpha_2$ domain we considered the
circular path of radius $0.15$ and centered in $(\alphac,\alphac)$, 
represented in figure \ref{fig_phdiagasymm}. 
In each one of $64$ equidistant points along this circle we determined
the stationary state
densities $\rho_1$ and $\rho_2$ as well as the currents $J_1$ and $J_2$
in the two lanes, for lane length $L=600$.
Each data point was obtained by first
discarding a transient period of $10\,000$ time steps
and then averaging over $100\,000$ time steps.
The results, together with the theoretical predictions,
are shown in figures \ref{fig_circleplotJ} and
\ref{fig_circleplotrho},
where $\phi$ is the angle between the radius vector on the circle and
the positive $\alpha_1$ axis.
Dotted vertical lines indicate the positions of the phase boundaries,
labeled by the same lettering $A,B,C,D$ as in figure \ref{fig_phdiagasymm}.
The error bars of the data points are smaller than the symbols.
The current data of figure \ref{fig_circleplotJ} fall perfectly on
the theoretical curve throughout the whole range of measurements.
The density data of figure \ref{fig_circleplotrho}
show a clear deviation from the theoretical 
prediction at those phase boundaries where the latter is discontinuous.
Indeed, the prediction is for lane length $L=\infty$
and these deviations appear to be finite size effects.
We have verified that indeed this effect decreases when $L$ goes up. 
Their physical origin is the same as was found for a single lane
\cite{appert-rolland_c_h11b}:
they result from the formation
of a fluctuating jammed boundary layer near the exit (when the lane is in
its free flow phase), or of a fluctuating free flow boundary layer 
near the entrance (when the system is in its jammed phase).


\section{Conclusion}
\label{sect_conclusion}

We have studied pedestrian traffic on two semi-infinite
one-dimensional lattices, or {\it lanes,}
that intersect in a common end point. 
The pedestrians are modeled as the particles of a TASEP, that is,
as hard core particles capable of moving only in a single direction,
in the present case toward the exit. 
When leaving the intersection site, a particle exits the system.
The particle positions were updated with `frozen shuffle' dynamics,
described in section \ref{sect_rules} and
argued to be a natural choice for pedestrian motion.
The updating is easy to implement in a Monte Carlo simulation an
also lends itself particularly well to analytical study.

Each of the lanes (labeled by $\sigma=1,2$)
is characterized by a parameter $\alpha_\sigma$ 
governing the entrance of the particles at $-\infty$,  
and another one,$\beta_\sigma$, governing 
their exit from the intersection site. 
For arbitrary fixed $\beta_1$ and $\beta_2$ 
we have determined analytically the phase boundaries in the
$\alpha_1\alpha_2$ plane.
It appears that each lane may be in either a free flow (F) or a jammed
(J) phase, which results in a partition of
the phase diagram in the $\alpha_1\alpha_2$ plane
into four regions, JJ, FJ, JF, and FF.
Explicit expressions have been found for 
the phase boundaries between these regions.
An essential element in our analysis is the
{\it pairing effect\,} that we have shown to occur when both lanes are in
the jammed phase: in that case each platoon exiting from one lane is
accompanied by a platoon exiting from the other lane.
Once the pairing effect is established,
the analytical expressions for the macroscopic
quantities of greatest interest, the currents and the particle
densities, become accessible via reasonably simple
mathematics.
We have determined them
analytically for each region of the phase diagram.
All our analytical findings have been corroborated by Monte Carlo
simulations presented in section \ref{sect_simulations}.

This work is to be seen as a first step towards
the study of the intersection of larger corridors,
in which case the possibility
of lateral hops may also have to be included. 
We will leave the study of such more complicated geometries and hopping rules
to future work.






\end{document}